\documentclass[journal]{IEEE}
\usepackage{tabularray}
\usepackage[colorlinks,urlcolor=blue,linkcolor=blue,citecolor=blue]{hyperref}
\usepackage{amsbsy}
\usepackage{color,array}
\usepackage{subcaption}  
\usepackage{graphicx}
\usepackage{multirow} 
\usepackage{cite}
\usepackage{svg} 
\usepackage{amsmath}
\usepackage{amssymb}
\usepackage{tabularx}  
\usepackage{adjustbox}
\usepackage{float}

\makeatletter

\newcommand{\Rmnum}[1]{\expandafter\@slowromancap\romannumeral #1@}
\makeatother

\begin{document}

\title{Lightweight CNN Model Hashing with Higher-Order Statistics and Chaotic Mapping for Piracy Detection and Tamper Localization}

\author{Kunming Yang, Ling Chen
\thanks{This work was supported by Chongqing Natural Science Foundation General Project under Grant CSTB2024NSCQ-MSX0428, in part by Chongqing Municipal Education Commission under Grant KJQN202300207, and also supported by project of State Key Laboratory of Integrated Chips and Systems under Grant No: SKLICS-G202503.}
\thanks{Kunming Yang is with the College of Electronic and Information Engineering, Southwest University, Chongqing 400715, China (e-mail: yang.kunming@foxmail.com).

Ling Chen is with the State Key Laboratory of Integrated Chips and Systems, Shanghai 201203, China, and the College of Electronic and Information Engineering, Southwest University, Chongqing 400715, China (e-mail: 2006chenling2006@163.com).
}
}


\maketitle
\begin{abstract}
With the widespread adoption of deep neural networks (DNNs), protecting intellectual property and detecting unauthorized tampering of models have become pressing challenges. Recently, Perceptual hashing has emerged as an effective approach for identifying pirated models. However, existing methods either rely on neural networks for feature extraction, demanding substantial training resources, or suffer from limited applicability and cannot be universally applied to all convolutional neural networks (CNNs). To address these limitations, we propose a lightweight CNN model hashing technique that integrates higher-order statistics (HOS) features with a chaotic mapping mechanism. Without requiring any auxiliary neural network training, our method enables efficient piracy detection and precise tampering localization. Specifically, we extract skewness, kurtosis, and structural features from the parameters of each network layer to construct a model hash that is both robust and discriminative. Additionally, we introduce chaotic mapping to amplify minor changes in model parameters by exploiting the sensitivity of chaotic systems to initial conditions, thereby facilitating accurate localization of tampered regions. Experimental results validate the effectiveness and practical value of the proposed method for model copyright protection and integrity verification.
\end{abstract}


\section{Introduction}

\IEEEPARstart{I}{n} recent years, deep neural networks (DNNs), particularly convolutional neural networks (CNNs), have achieved remarkable success across a wide range of fields, including agriculture \cite{ref1, ref2}, industrial inspection \cite{ref3, ref4, ref5}, and natural language processing \cite{ref6, ref7}. The training of such models typically relies on substantial computational resources, large-scale labeled datasets, and extensive iterative optimization. As a result, a high-performing neural network model often encapsulates valuable intellectual property (IP) and carries significant competitive advantages. However, during the deployment and distribution phases, models are highly susceptible to unauthorized duplication, dissemination, and malicious tampering. Unauthorized copying and distribution can directly infringe upon the IP rights of model owners, while deliberate manipulation of model parameters may lead to degraded performance in critical applications or even introduce hidden backdoors.

These challenges highlight the urgent need for effective techniques in model piracy detection and tampering localization. Robust model hash and tamper detection mechanisms are not only essential for safeguarding the IP of deep learning models, but also for ensuring the integrity and trustworthiness of artificial intelligence systems in security-sensitive scenarios. Currently, research on neural networks copyright protection and integrity verification primarily falls into three categories: model watermarking, model fingerprinting, and perceptual hashing. Model watermarking techniques \cite{ref8, ref9} embed specific ownership information into model parameters, allowing legitimate owners to extract this information as proof of ownership. However, such approaches often rely on directly modifying model weights or retraining the models, which may risk performance degradation or disrupt existing deployment pipelines.

To avoid directly modifying model parameters, researchers have proposed model fingerprinting techniques \cite{ref10, ref11}. These approaches typically verify model ownership by constructing adversarial queries or utilizing auxiliary data to probe the model’s behavior. Although such methods provide an effective verification mechanism without altering the model’s structure, they share a common limitation with watermarking approaches in practical applications: when faced with a large-scale model repository, each candidate model must be verified individually, resulting in inefficiencies that make large-scale retrieval infeasible.

Over the past few years, perceptual hashing has gained rapid traction in the field of deep learning model protection. Its core concept is inspired by feature retrieval techniques used in multimedia content security \cite{ref12, ref13, ref14}, and it offers distinct advantages. By encoding a model’s essential features into compact hash representations and storing them centrally, this method enables fast similarity comparisons against suspicious models, thereby facilitating rapid piracy detection and integrity verification. For example, the method proposed by Chen et al. \cite{ref15} employs normal test statistics (NTS) to extract model features and generate hash codes. However, this method assumes the model is trained with L2 regularization and specific initialization schemes. When such assumptions are violated, the method’s generalizability is significantly compromised. Furthermore, this approach does not support tampering localization. Fig. 1 illustrates the limitations of NTS-based method.

\begin{figure}
\centerline{\includegraphics[width=21pc]{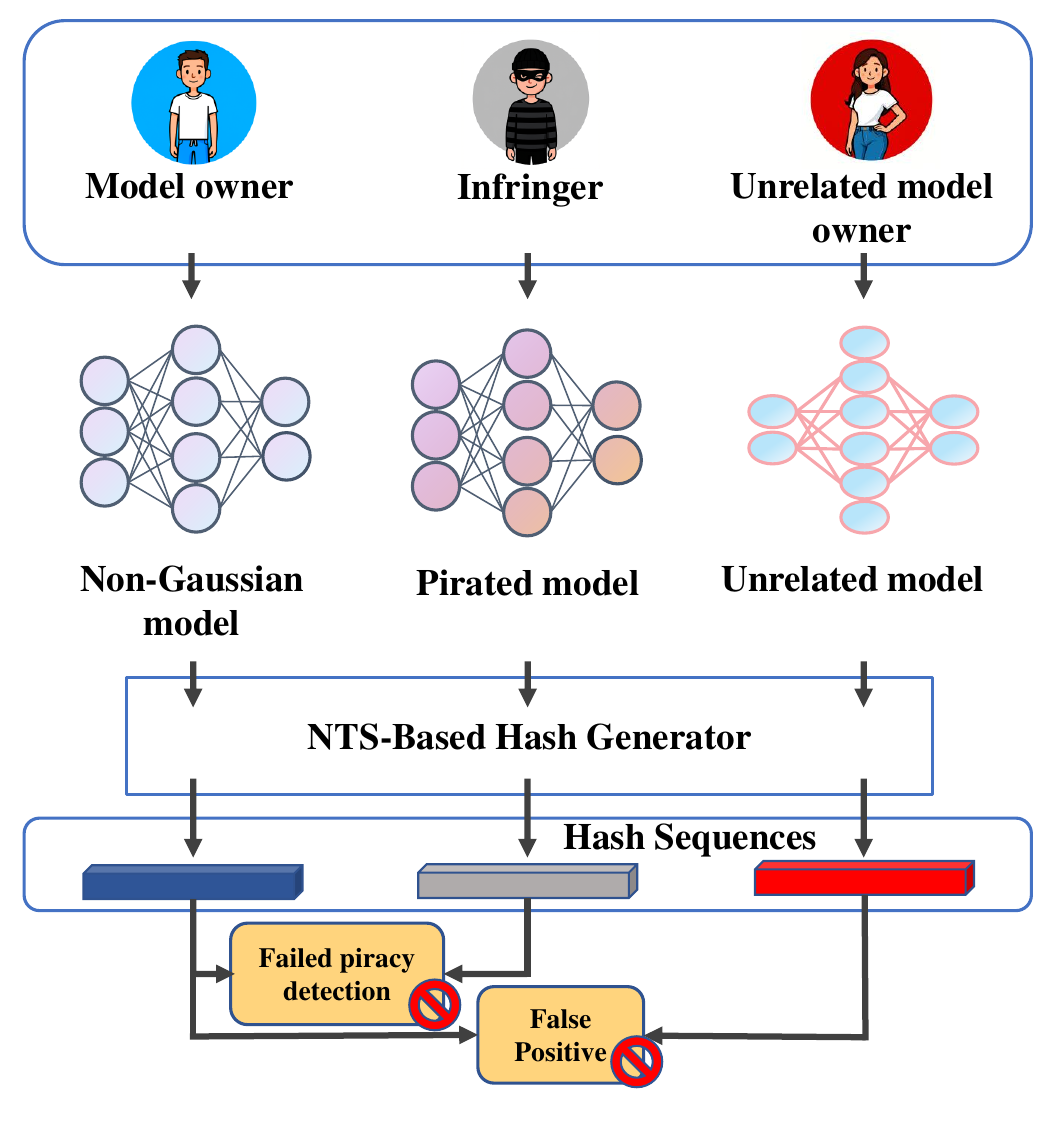}}
\caption{An illustration of limitations inherent to NTS-based method.}
\end{figure}
In contrast, hashing frameworks based on DNNs exhibit stronger adaptability, such as those proposed by Xiong et al. \cite{ref16} and Chen et al. \cite{ref17}. These methods not only identify model ownership but also localize specific regions where parameters have been tampered with. Nevertheless, they require extensive training on large numbers of labeled model pairs (e.g., "pirated vs. original"), incurring substantial time and computational overhead.

To address the limitations of existing techniques in terms of scalability and resource demands, we propose a lightweight CNN model hashing method that combines higher-order statistics (HOS) features with the chaotic mapping mechanism. Unlike prior approaches, our method does not rely on additional deep learning model training nor on assumptions about parameter distributions, making it highly efficient for piracy detection and accurate tampering localization. The core idea is to extract discriminative features such as skewness, kurtosis, and structural information from each network layer, so as to generate a hash representation that is both robust and distinctive. In addition, we innovatively introduce chaotic mapping, which exploits the extreme sensitivity of chaotic systems to initial conditions. This allows for the amplification of minor perturbations in model parameters, enabling precise localization of tampered regions.

Experimental results demonstrate that the proposed method can reliably identify protected models under various modifications, including pruning, fine-tuning, and knowledge distillation, while effectively distinguishing them from unrelated models. In tampering detection tasks, the method achieves an average localization accuracy of up to 99.9$\%$, confirming its strong potential for applications in CNNs copyright protection and integrity verification.

The main contributions of this study are as follows:

(1) We propose a feature extraction framework that integrates HOS features with model structural information, thereby broadening the applicability of model hashing techniques. This framework enables effective identification of pirated models and precise discrimination from unrelated ones.

(2) We introduce the chaotic mapping mechanism for model tampering detection, along with a chaotic enhancement strategy that substantially improves sensitivity to subtle parameter manipulations. The method achieves an average localization accuracy exceeding 99.9$\%$ across various tampering ratios.

(3) Without relying on any neural network training, we develop a lightweight solution for model piracy detection and tampering localization using only interpretable features such as HOS and model structure.

\section{RELATED WORK}

\subsection{Model Hashing}

Research on copyright protection for DNN models can be traced back to the model watermarking method proposed by Uchida et al. \cite{ref8} This approach embeds ownership-identifiable watermarks into model weights by incorporating a parameter regularization mechanism during training, enabling verifiable ownership claims without significantly degrading model performance. To avoid directly modifying model parameters, Zhao et al. \cite{ref10} further introduced the adversarial fingerprinting authentication method. This technique verifies ownership by generating carefully crafted queries that probe the model's inherent features. However, fingerprinting techniques typically rely on targeted testing and are highly model-dependent, which makes them difficult to apply in large-scale retrieval or fast matching scenarios. To balance retrieval efficiency and method generalizability, Chen et al. \cite{ref15} proposed a model hashing method for CNNs inspired by perceptual hashing techniques from image processing. Their method extracts features from network weights and maps them into a fixed-length binary hash code. This hash remains consistent under lightweight model modifications, such as pruning or fine-tuning, thus allowing fast similarity comparison based on hash distance.

However, the NTS-based method is built upon a crucial assumption: the model must be trained using L2 regularization and a specific initialization scheme, which together are expected to yield approximately Gaussian-distributed weights \cite{ref18, ref19}. Under this assumption, the deviation of the model weights from a Gaussian distribution is used as the basis for hash construction. In practice, however, such assumptions do not always hold, especially when alternative regularization strategies or initialization methods are employed, which may lead to markedly non-Gaussian parameter distributions that can undermine the method’s stability and applicability.

Subsequently, Xiong et al. \cite{ref16} introduced a dual-branch network architecture to replace traditional statistical feature extraction and manual quantization processes. They also incorporated a tampering localization mechanism to detect and indicate specific regions of the model parameters that had been modified. However, the high false positive rate of this approach poses challenges for practical applications. To further enhance robustness and applicability, Chen et al. \cite{ref17} proposed a method that converts various neural network architectures into graph representations and utilizes graph neural networks (GNNs) for feature learning. This method reduced the false positive rate for non-pirated models to 3$\%$, while achieving a detection accuracy of over 98$\%$ for pirated models.

Despite these advances in model hashing, particularly in automating feature extraction and supporting tampering detection, these approaches share a common limitation: their reliance on DNNs, which requires large datasets and substantial training resources. This significantly increases the deployment threshold and limits their practicality in resource-constrained environments, such as embedded systems and edge computing scenarios.

To address these issues, we propose a lightweight model hashing method based on HOS features, capable of efficiently detecting model piracy without requiring additional neural network training. In addition, we incorporate a chaotic mapping mechanism, which, due to its sensitivity to initial conditions, amplifies minor variations in model weights. This significantly improves the detection and localization of tampering behavior.

\subsection{Higher-order Statistics}

HOS originated from the study of signal distribution characteristics and are primarily used to describe features of probability distributions beyond the mean and variance. Classical statistics typically relies on lower-order moments, such as the first-order moment (mean) and second-order moment (variance), to describe the central tendency and dispersion of data. In contrast, HOS include skewness, kurtosis, and higher-order cumulants, which can further reveal the asymmetry, peakedness, and higher-order dependencies within probability distributions.

HOS has been widely applied in fields such as radar signal processing, speech recognition, and seismic data analysis, largely due to its inherent robustness against Gaussian noise. Unlike lower-order features that depend solely on variance, HOS can detect non-Gaussian characteristics in signals under Gaussian interference, thereby enabling the extraction of richer and more robust feature information.

For instance, Barreira et al. \cite{ref20} addressed the pronounced non-Gaussianity in the distribution of cosmic matter by employing HOS, thereby overcoming the limitation of the power spectrum, which is derived from second-order statistics, in capturing the full range of cosmological information. This methodology enabled tighter constraints on key cosmological parameters, including the dark energy density and neutrino mass. Similarly, Khoshnevis et al. \cite{ref21} applied HOS analysis to resting-state electroencephalography (EEG) data and demonstrated that HOS-derived metrics can reliably distinguish patients with Parkinson’s disease from healthy controls. Their findings provide compelling evidence that HOS are well-suited to characterizing the nonlinear and non-Gaussian features of brain activity. Collectively, these studies highlight the broad utility of HOS as a powerful tool for feature extraction across diverse scientific domains.
\subsection{Chaotic Mapping}

Chaotic mapping refers to a class of discrete-time nonlinear dynamical systems characterized by extreme sensitivity to initial conditions. Specifically, even a minute perturbation in the initial state of a chaotic system can be exponentially amplified during the iterative evolution process, leading to drastically different output sequences. This property results in highly random and unpredictable behavior.

Since the 1980s, chaos theory has been widely applied in fields such as secure communications, pseudo-random number generation, information encryption, and digital watermarking. Typical forms of chaotic maps include the Logistic map, Tent map, and Henon map. These systems exhibit complex dynamical behavior, strong ergodicity, and inherent irreversibility, providing a solid theoretical foundation for security-critical applications.

In the context of information security, chaotic mapping has been shown to enhance the concealment and tamper resistance of sensitive data. For instance, chaotic sequences are commonly used for key stream generation and image encryption. In this work, we apply chaotic systems to model tampering detection by leveraging their sensitivity to initial conditions. This allows for the amplification of subtle changes in model parameters during the hashing process, thereby enabling accurate localization and effective identification of tampering behavior.

In particular, we adopt a second-order memristive chaotic map \cite{ref22} as the core chaotic mechanism in our framework. This system is structurally simple yet exhibits high chaotic complexity and hyperchaotic behavior, making it well-suited for precise tampering localization and protection of neural network models.

\section{Problem Formulation}

Let us consider a CNN model $M$, the core objective of model hashing is to design a function $f$ that generates a fixed-length binary hash code for model $M$, which can be used for piracy detection and tampering localization:
\begin{equation}
f(M,key)=H,H\in\{0,1\}^t
\end{equation}
where $H$ is a binary hash sequence of length $t$, and $key$ is a secret key.

The hash function constructed in this work is expected to satisfy the following three properties:

(1) Discriminability: Hash codes corresponding to different models should exhibit significant differences.

(2) Robustness: Hash codes should remain relatively stable when the model undergoes typical benign modifications such as pruning, fine-tuning, or knowledge distillation.

(3) Sensitivity: When a model is maliciously tampered with, even minimal changes to its parameters should be reflected in the hash code, allowing accurate localization.

Formally:
If $M_a$ is a pirated copy of $M_b$, then $D(H_a,H_b)<\tau$.

If $M_a$ and $M_b$ do not belong to the same IP, then $D(H_a,H_b)>\tau$.

If $M_a$ is a tampered version of $M_b$, then $D(H_a,H_b)>0$.

Here, $M_a$ and $M_b$ represent two models to be tested, $D$ represents the hash distance, and the threshold $\tau$ represents the tolerance range for slight modifications to the model (such as fine-tuning, pruning, and knowledge distillation).

\section{Proposed Method}

In this section, we systematically elaborate on the overall framework of the proposed perceptual hashing algorithm. Our piracy identification hash first selects important weights based on L1 pruning to reduce redundancy and enhance feature stability. Subsequently, the filtered weight sequence is partitioned into several non-overlapping segments. The skewness and kurtosis of each segment are computed respectively to generate a fixed-length HOS feature sequence. Following this, we generate a fixed-length network structural feature sequence based on the model's architectural characteristics. Finally, through processes such as quantization and encryption, these HOS features and structural feature sequences are mapped into a compact binary hash code. By comparing the hash similarity between the target query model and candidate models, potential model copies can be efficiently identified. For the tamper localization hash, model parameters are first divided into a predetermined number of parameter blocks. The mean value of each block is computed and then processed through the chaotic mapping system. The system's output is finally converted into the tamper localization hash via a custom binary encoding rule. The following subsections provide detailed explanations of each component of the proposed method.

\subsection{Piracy Identification Hash}
\subsubsection{HOS Feature Sequence}
(1) Weight Selection. Given the significant redundancy in CNNs weights, inspired by model compression theory, we employ the L1 norm of weights to measure their importance, thereby implementing L1-norm-based weight selection. Specifically, the hash code is constructed from weights with larger absolute values. Formally, let $\mathbf{v}$ be the list of absolute values of all weights in model $M$, and $c$ be the target proportion of weights to retain (referred to as the weight selection ratio in this work). We calculate a threshold $th_{\mathbf{v}}$ such that the proportion of weight magnitudes below this threshold equals $q = 1-c$:
\begin{equation}
th_{\mathbf{v}}=quantile(\mathbf{v},q)
\end{equation}
where $quantile(\cdot,q)$ denotes the function computing the $q$-th quantile of a list.

For the list $\mathbf{v}$ = [$v_{1},v_{2},\ldots,v_{n}$], we first sort it in ascending order to obtain $\mathbf{v^\prime}$ = [$v_{(1)},v_{(2)},\ldots,v_{(n)}$], then its $q$-th quantile can be calculated as:
\begin{equation}
Q_q=(1-d)\cdot v_{(k+1)}+d\cdot v_{(k+2)}
\end{equation}
where $Q_q$ represents the $q$-th quantile of list $\mathbf{v}$, $d=r-k$, $k=\lfloor r\rfloor$, and $r=q\cdot(n-1)$. Subsequently, weights with magnitudes smaller than $th_{\mathbf{v}}$ are excluded from hash code computation.

(2)Fixed-length Feature Generation. In this subsection, we compute the kurtosis and skewness of pruned weight segments to obtain robust features. Skewness measures the asymmetry of a random variable's probability distribution, while kurtosis quantifies the distribution's peakedness.

As previously discussed, to ensure the robustness of hash features, key characteristics should be extracted from weights that significantly impact the overall performance of DNNs. The scheme proposed in \cite{ref15} assumes that "weight importance can be measured by its distance from a Gaussian distribution" \cite{ref19}. This assumption primarily stems from the common use of L2-norm regularization during deep learning training to prevent overfitting, which tends to make CNNs filter parameters approximately Gaussian-distributed \cite{ref5}. However, when uniform initialization replaces Kaiming initialization, or when L2 regularization is omitted during training, model parameters typically deviate from Gaussian distribution. In such cases, feature extraction methods based on normality tests lose effectiveness.

\begin{figure}
    \centering  
    \begin{subfigure}[b]{21pc}
        \centering
        \includegraphics[width=\textwidth]{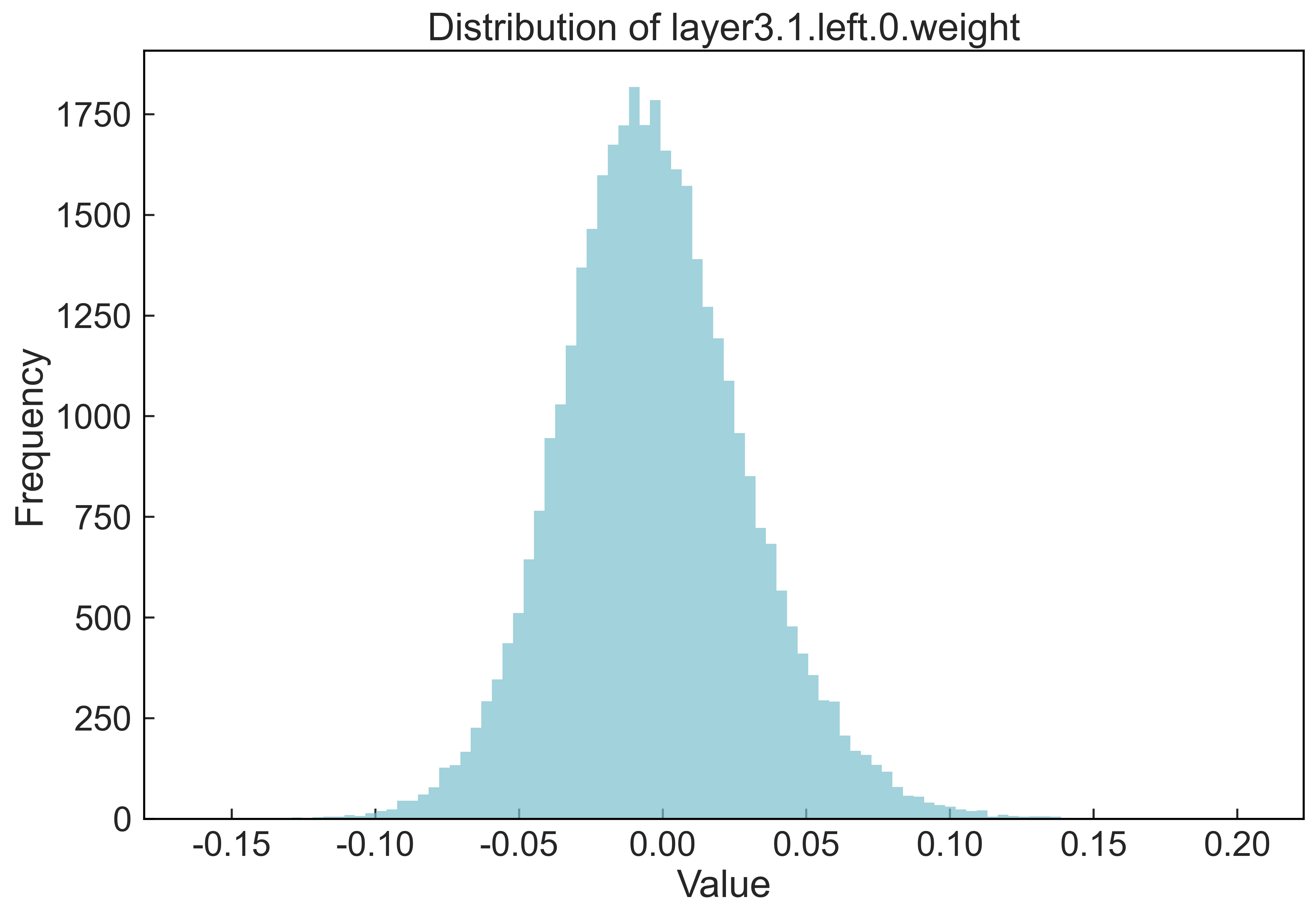}
        \caption{L2 regularization and Kaiming initialization}  
    \end{subfigure}
    \hfill  
    \begin{subfigure}[b]{21pc}
        \centering
        \includegraphics[width=\textwidth]{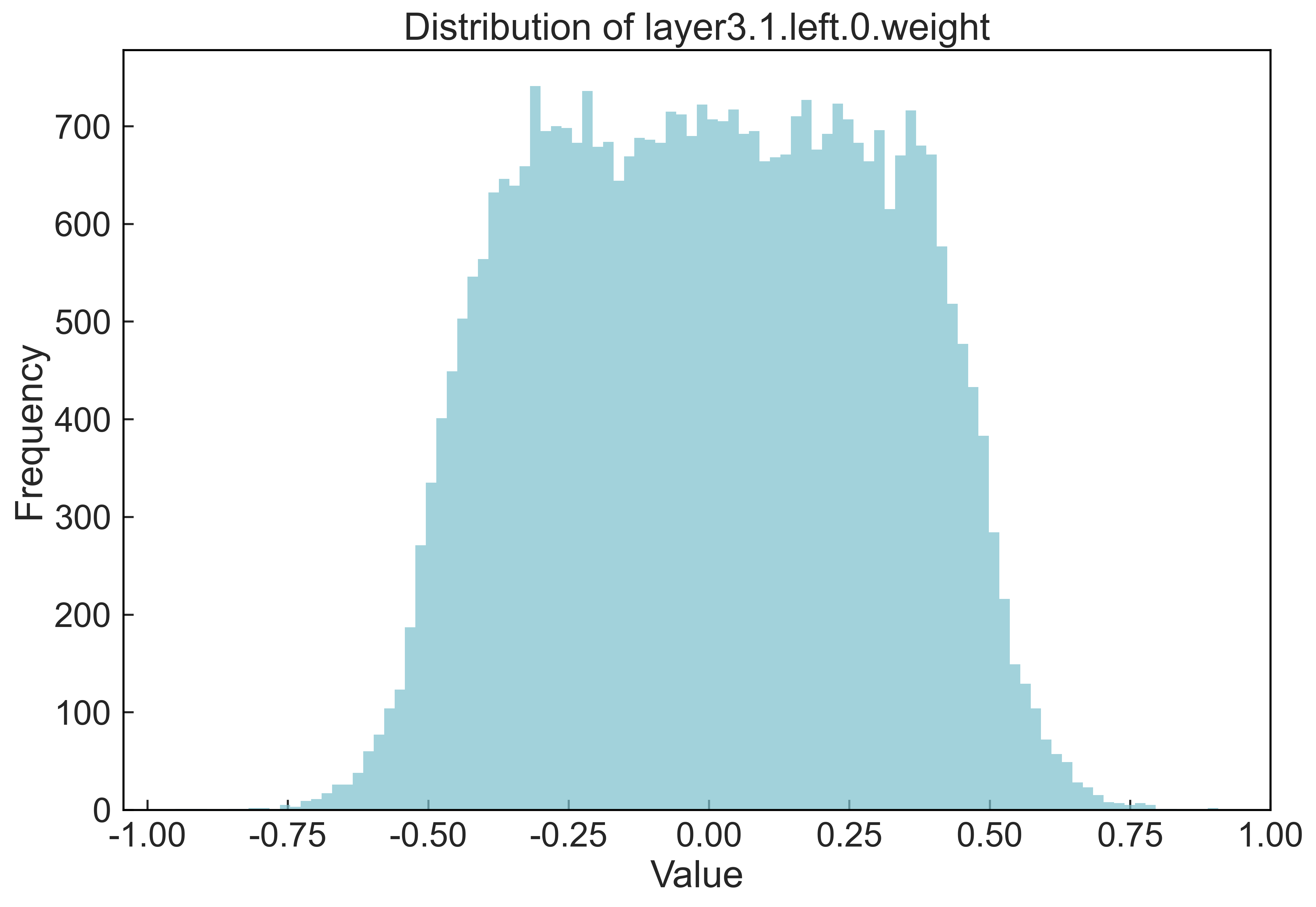}  
        \caption{Uniform initialization without regularization}
    \end{subfigure}
    \caption{Comparison of parameter distributions between ResNet32 models trained with L2 regularization and Kaiming initialization versus uniform initialization without regularization.}  
    \label{fig:para}  
\end{figure}
For example, Fig. 2 presents a comparison of the parameter distributions derived from models trained using L2 regularization in conjunction with Kaiming initialization versus those trained with uniform initialization without regularization. As can be observed, the use of uniform initialization results in parameter distributions across the network layers that deviate from a Gaussian distribution. This deviation is inconsistent with the prerequisite of "weight parameters following an approximately Gaussian distribution," which is necessary for employing the NTS-based method to extract model parameters features.

In contrast, our proposed method does not rely on any Gaussian distribution prior. Instead, it directly utilizes HOS like skewness and kurtosis to characterize weight distribution properties. This approach ensures stable extraction of discriminative features even when parameter distributions deviate from Gaussian assumptions, thereby enhancing the generalizability and robustness of model hashing.

Specifically, for model $M$, whose set of parameters is denoted as $\mathcal{W}=\{W_l\}_{l=1}^L$, where $L$ represents the total number of layers, and $W_l$ denotes the weight tensor of the $l-th$ layer, we flatten the weights of each layer into a 1D array $r_l$, resulting in a flattened weight sequence $R=\{r_1,r_2,...,r_L\}$ after processing all layers. We then equally divide the concatenated vector $R$ into $N$ non-overlapping segments $R^{\prime}=\{r_1^{\prime},r_2^{\prime},...,r_N^{\prime}\}$. For each segment $r_i^{\prime}$, we calculate its skewness $s_i$ and kurtosis $k_i$ as features representing the model's perceptual content. This yields an HOS sequence $\mathbf{a}=\{s_{1},s_{2},...,s_{N},k_{1},k_{2},...,k_{N}\}$ for the $N$ segments. Given $r_{i}^{\prime}=[x_{1},\ldots,x_{n}]$ where $x_i$ represents weights, skewness ($S$) and kurtosis ($K$) are computed as:

\begin{equation}
S=\frac{1}{n}\sum_{i=1}^n[(\frac{X_i-\mu}{\sigma})^3]
\end{equation}

\begin{equation}
K=\frac{1}{n}\sum_{i=1}^n[(\frac{X_i-\mu}{\sigma})^4]
\end{equation}
where $\sigma$ represents standard deviation and $\mu$ denotes the mean.
\subsubsection{Network Structure Feature Sequence}
HOS demonstrate clear advantages in robustness compared to NTS, but its discriminative capability for distinguishing different network architectures remains relatively limited. Therefore, we proposes incorporating model structure-aware features alongside HOS to construct a composite hash, thereby achieving both robustness and discriminability simultaneously.

Specifically, we first calculate the total trainable parameters for each convolutional layer and count the total number of convolutional layers. This count serves as a coarse-grained structural descriptor representing network depth and complexity. Next, we normalize the parameter count of each convolutional layer by dividing each layer's parameter quantity by the sum of all convolutional layer parameters, yielding a proportional vector that reflects each layer's contribution to the overall model capacity. To ensure dimensional consistency across different network architectures, we apply uniform processing to the normalized proportional vector: when the number of convolutional layers exceeds a preset hyperparameter $K$=20, we retain the first $K$ features; when the count is insufficient, we perform zero-padding to complete the dimensionality.

Finally, we use the normalized layer count (i.e., the ratio of actual convolutional layers to $K$, capped at 1.0) as the first element, concatenated with the normalized parameter proportion vector, to generate the final structural feature sequence with dimensionality 1+$K$.

Mathematically, if a model contains $P\leq K$ convolutional layers with parameter counts $\{p_{1},p_{2},\ldots,p_{P}\}$ and total parameter count $P_{Num}=\sum_{i=1}^{P}p_{i}$, the structural feature sequence $\mathbf{b}$ is defined as:
\begin{equation}
\mathbf{b}=\left[min\left(\frac{P}{K},1.0\right),\frac{p_1}{P_{Num}},...,\frac{p_P}{P_{Num}},0,...,0\right]
\end{equation}

This sequence compactly encodes key architectural information, including the total convolutional layer count and each layer's parameter proportion. 
\subsubsection{Encoding}
In the encoding stage, the HOS feature sequence and network structure feature sequence are transformed into a compact binary representation. Specifically, the sequence is first stabilized by offsetting and applying a logarithmic scaling to compress its dynamic range. The log-transformed values are then normalized to the interval [0,1], ensuring consistency across different sequences. Each normalized statistic is uniformly quantized into $2^b$ discrete levels, where $b$ denotes the number of bits assigned per statistic. Finally, the quantized values are converted into fixed-length binary strings and concatenated to form the binary vector $\boldsymbol{\gamma}$. This process effectively maps HOS and network structure features into a discriminative and storage-efficient binary space, facilitating subsequent fast similarity measurement and feature matching. By combining the HOS features with the network structure features, we obtain the binary vector, where structural features enhance discriminability while HOS features maintains robustness.

Security requirements dictate that hash codes must be unestimable without knowledge of the secret key. Otherwise, adversaries could potentially estimate hash codes to fine-tune duplicated models, creating adjusted models that maintain similar performance to the original but exhibit significantly different hash values, thereby evading copy detection. Therefore, the final piracy identification hash $H_p$ is encrypted using a secret key through the following formula to prevent malicious circumvention:
\begin{equation}
H_p=\boldsymbol{\gamma}\oplus key
\end{equation}
where $key$ represents pseudorandom bits generated from the secret key, and $\oplus$ denotes the bitwise XOR operation.
\subsection{Tampering Localization Hash}
To accomplish the model tampering localization task, we first load the trained model parameters and transform the multi-dimensional weight tensors into a consistent one-dimensional vector representation through flattening and normalization. Subsequently, we partition the normalized weights into a predetermined number of parameter blocks and compute the mean value of weights within each block as local features. Finally, we feed these N extracted local features into the discrete second-order memristive chaotic system in (8), run it for a fixed number of iterations (100 iterations in this work), and convert the resulting $N$ chaotic feature values into tamper detection hash codes using a custom binary encoding scheme. These codes are then concatenated to form the final tamper localization hash.
\begin{equation}
	\begin{split}
x_{n+1} & =\mu x_{n}\left(1-x_{n}\right)+k \cos q_{n} \cdot x_{n} \\
q_{n+1} & =q_{n}+x_{n}
	\end{split}
\end{equation}

For each parameter block, $x_0$ is initialized as the mean value of weights within the block, with $q_0=x_0/2$, $\mu=0.2$, and $k=2$. As shown in the Fig. 3, both Lyapunov Exponents (LEs) of the chaotic system are positive under these parameters, indicating the system operates in a hyperchaotic state with high complexity.
\begin{figure}
\centerline{\includegraphics[width=21pc]{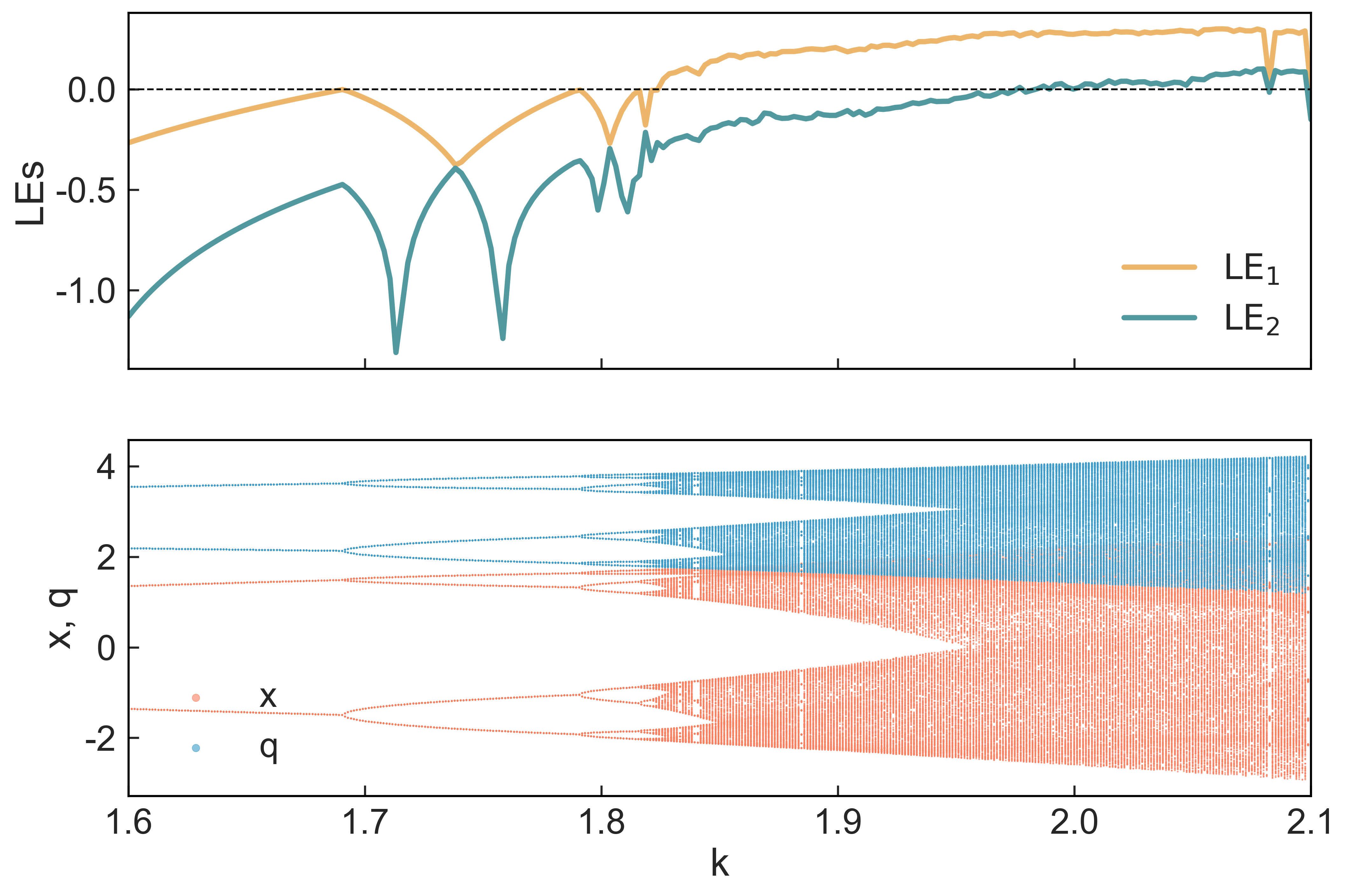}}
\caption{For the two-dimensional system, the bifurcation diagrams of the LEs (top) and the state variables $x$ and $q$ (bottom) as functions of the parameter $k$.}
\end{figure}

To enable efficient storage and comparison of chaotic values, this study designs a customized fixed-length binary encoding scheme. This approach combines the extraction of significant decimal digits with fixed-width per-digit binary encoding for model integrity verification.

Specifically, the encoding process first separates the sign component and truncated magnitude from each chaotic feature value generated by the chaotic system. The sign component is encoded using a single-bit binary indicator (0/1) to represent the positivity/negativity of the feature value, while the truncated magnitude is obtained by preserving only the first few significant digits (the first 4 digits in our implementation) of the decimal value. Each digit of the truncated magnitude is then converted into a fixed-width (4 bits in our implementation) binary representation. These binary segments are concatenated with the sign bit to form the final tamper localization hash code with uniform length after undergoing the same encryption process as the piracy identification hash.

This encoding strategy achieves both compactness and high discriminability. Even minimal perturbations in input feature values produce substantially different binary encodings, thereby providing precise support for model integrity verification and tamper detection tasks. Ultimately, by comparing the tamper localization hashes generated before and after model modification, we can accurately identify and locate the specific parameter blocks that have been tampered with.
\section{Experiments}
\subsection{Setting}
This section first details the experimental setup, then systematically evaluates the proposed perceptual hashing algorithm across four key dimensions: discrimination capability, robustness, tamper localization accuracy, and computational cost.

\subsubsection{Model}
To comprehensively assess the adaptability and effectiveness of our method under various conditions—particularly its robustness and discriminative power when model parameters deviate from Gaussian distributions, We compared our approach with existing NTS-based perceptual hashing methods in \cite{ref15}, employing a "uniform initialization + no regularization" training strategy to generate test models.

As specified in Table \Rmnum{1}, we select eight widely-used convolutional neural network architectures for evaluation. Specifically, the first six models were trained on the CIFAR-10 dataset \cite{cifar}, while MNIST-Net and LeNet were trained on the MNIST dataset \cite{MNIST}.

\begin{table*}[h]
    \centering
    \caption{The structure of models and datasets.}\label{tab1}
    \adjustbox{max width=\textwidth}{ 
    \begin{tabular}{c|c|c|c|c|c|c|c|c}
        \hline
        Model name & ResNet18 \cite{res18-32-56-110} & ResNet32 \cite{res18-32-56-110} & ResNet56 \cite{res18-32-56-110} & ResNet110 \cite{res18-32-56-110} & DenseNet121 \cite{densenet} & VGG16 \cite{vgg16} & MNIST-Net \cite{mnist-net} & LeNet \cite{lenet} \\
        \hline
        Dataset & CIFAR-10 & CIFAR-10 & CIFAR-10 & CIFAR-10 & CIFAR-10 & CIFAR-10 & MNIST & MNIST \\
        \hline
        Accuracy (\%) & 91.8 & 92.6 & 92.7 & 93.3 & 92.0 & 91.4 & 94.7 & 89.6 \\
        \hline
    \end{tabular}
    }
\end{table*}

\subsubsection{Metric}

To quantify the similarity between hash codes, we employ the Hamming distance as our evaluation metric:
\begin{equation}
D(H_{1},H_{2})=\frac{1}{t}\sum_{i=1}^{t}|H_{1}(i)-H_{2}(i)|
\end{equation}

To balance the robustness contributed by HOS hashes and the discriminability from network structural hashes, we introduce control parameters $k_1$ and $k_2$ (In this work, $k_1$ = 0.8 and $k_2$ = 0.2 are adopted.) to weight their respective contributions:
\begin{equation}
D(H_1,H_2)=k_1*D(h_1,h_2)+k_2*D(h_3,h_4)
\end{equation}
where $h_1$ and $h_2$ represent HOS hash codes, while $h_3$ and $h_4$ denote network structure hash codes. $H_1$ and $H_2$ are piracy identification hash codes derived from the concatenation of the aforementioned two types of hash codes. In the event that $D(H_1,H_2)$ is less than a predefined threshold, the CNNs corresponding to the input hash codes are deemed to be similar models.

For evaluating tamper localization performance, we compare the tamper localization hashes between original and modified models. The localization accuracy $R _t$ is formally defined as:
\begin{equation}
R_t=\eta^{\prime}/\eta
\end{equation}
where $\eta$ represents actual number of tampered parameter blocks and $\eta^{\prime}$ denotes the number of correctly detected tampered blocks.

\subsubsection{Implementation Details}

To ensure fair comparability with the NTS-based method in \cite{ref15}, our implementation strictly adheres to its parameter settings: the hash threshold $\tau$ is maintained at 0.32 and the weight selection ratio $c$ at $1/16$. Since we have modified the generation method of hash codes, the length of the hash code for piracy identification in this study is set to $T$ = 484. To control variables, the length of the NTS-based hash sequence generated in this section is also 484.
\subsection{Discrimination}
This evaluation examines the method's capacity to generate structurally discriminative hash representations when model parameters violate Gaussian distribution assumptions. We conduct comprehensive testing across all eight CNN architectures in Table \Rmnum{1}, generating unique hash codes for each model and computing pairwise Hamming distances.

Table \Rmnum{2} reveals two fundamental characteristics: all diagonal entries representing self-comparisons exhibit values approaching zero, while off-diagonal entries reflecting cross-model comparisons consistently exceed the 0.32 decision threshold. This distinct separation between near-zero intra-model distances and substantially higher inter-model distances demonstrates robust discrimination capability in non-Gaussian parameter environments. The consistent pattern across all tested architectures confirms the method's effectiveness in distinguishing fundamentally different models regardless of distributional assumptions.

\begin{table*}[h]
    \centering
    \caption{Discrimination on models with different structures.}\label{tab2}
    \begin{tabularx}{\textwidth}{>{\centering\arraybackslash}X|*{8}{>{\centering\arraybackslash}X}}
    \hline
              & ResNet32 & VGG16  & ResNet18 & ResNet56 & ResNet110 & DenseNet121 & LeNet  & MNIST-Net \\ \hline
    ResNet32    & 0        & 0.3876 & 0.4521   & 0.3885   & 0.3906    & 0.4006      & 0.4416 & 0.4184    \\
    VGG16       & 0.3876   & 0      & 0.4795   & 0.3908   & 0.3631    & 0.3937      & 0.3485 & 0.3622    \\
    ResNet18    & 0.4521   & 0.4795 & 0        & 0.4417   & 0.4309    & 0.4534      & 0.4815 & 0.5015    \\
    ResNet56    & 0.3885   & 0.3908 & 0.4417   & 0        & 0.3731    & 0.3692      & 0.3960 & 0.4528    \\
    ResNet110   & 0.3906   & 0.3631 & 0.4309   & 0.3731   & 0         & 0.3696      & 0.3684 & 0.3651    \\
    DenseNet121 & 0.4006   & 0.3937 & 0.4534   & 0.3692   & 0.3696    & 0           & 0.5085 & 0.3925    \\
    LeNet       & 0.4416   & 0.3485 & 0.4815   & 0.3960   & 0.3684    & 0.5085      & 0      & 0.3923    \\
    MNIST-Net   & 0.4184   & 0.3622 & 0.5015   & 0.4528   & 0.3651    & 0.3925      & 0.3923 & 0         \\ \hline
    \end{tabularx}
\end{table*}

\subsection{Robustness}
It should be pointed out that a large distance between hash codes extracted from irrelevant models does not necessarily mean that the hashing system is reliable. If the system cannot effectively identify the identity between the original model and its versions after common modifications, it may lead to an excessively high false positive rate, affecting the discrimination accuracy in practical applications.

To systematically evaluate the robustness and applicability of the proposed method in the face of model perturbations, this study constructs a series of typical perturbation scenarios, including model fine-tuning, pruning, and knowledge distillation, and compares the original model with the versions modified through the above methods respectively. Specifically, we use the three mentioned methods to generate these modified CNNs versions. The verification process starts with extracting the hash codes of the original CNNs and their similar versions, and calculates the similarity between each pair of hash codes through the normalized Hamming distance. Theoretically, the normalized Hamming distance between the original model and its similar versions should be less than the threshold of the hash codes.

\subsubsection{Fine-tuning}
Model fine-tuning is a common evasive attack method, where attackers attempt to evade hash retrieval by adjusting model weights. In the experiments of this work, the SVHN dataset is used to retrain the model trained on CIFAR-10, so as to alter the parameter distribution. Since MNIST-Net and LeNet are not applicable to the SVHN dataset, these two models are excluded from the fine-tuning experiments.

Fig. 4 shows the normalized Hamming distances between the models and their fine-tuned versions under different numbers of fine-tuning epochs, with the horizontal axis representing the number of fine-tuning iterations and the vertical axis representing the normalized distance. From the results, it can be observed that all normalized Hamming distances obtained by the HOS-based method are below the set threshold, indicating that this method can stably generate consistent hash codes and effectively identify similar models under the condition that the parameter distribution is non-Gaussian. In contrast, the distances of the NTS-based method are all above the threshold, indicating that it lacks robustness against fine-tuning.
\begin{figure}
\centerline{\includegraphics[width=21pc]{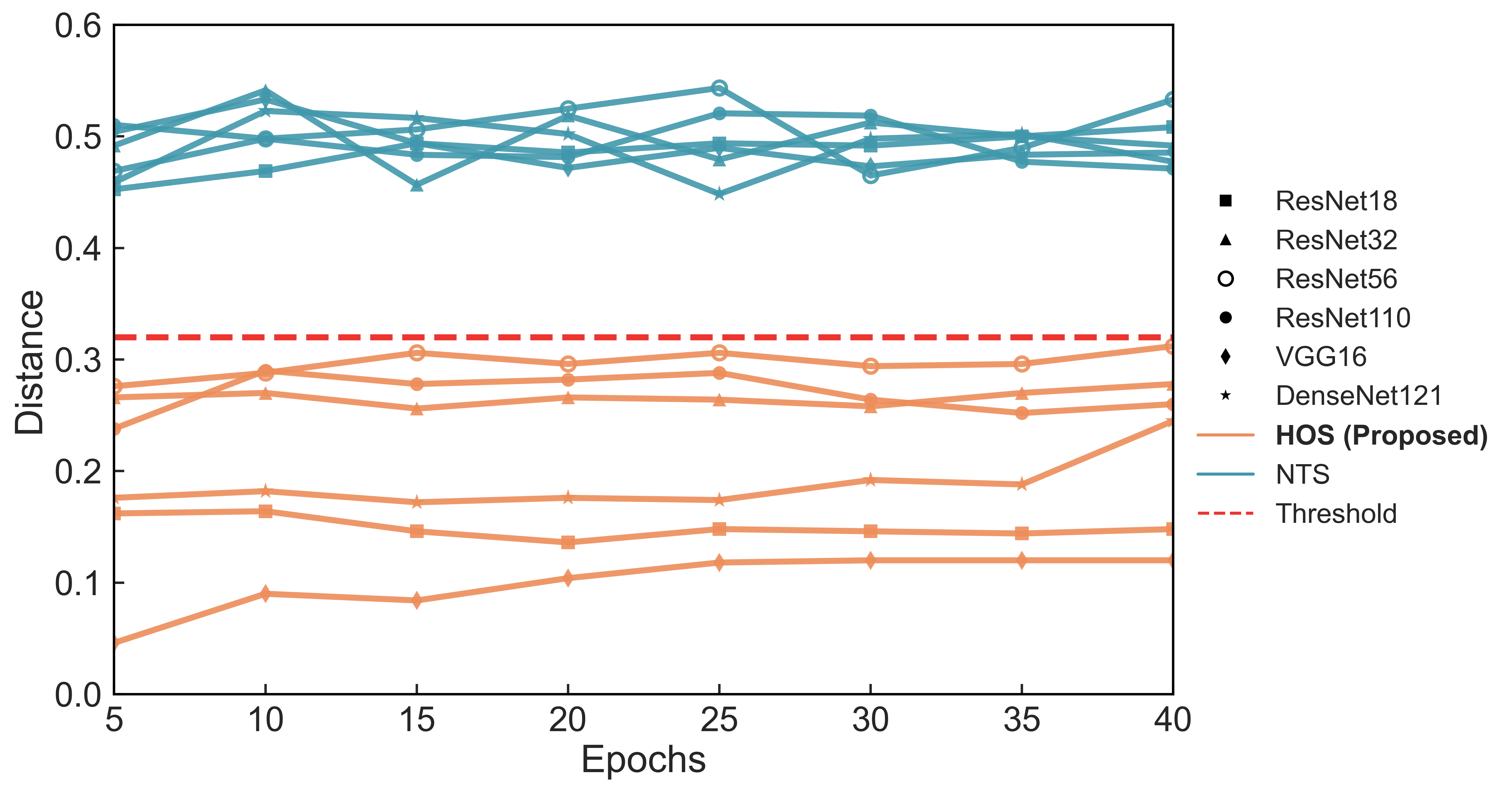}}
\caption{Normalized hamming distances of hash codes between the original models and the fine-tuned versions using NTS and HOS methods.}
\end{figure}

\subsubsection{Pruning}
It is assumed that attackers sparsify the model through a magnitude-based pruning strategy to avoid retrieval. In the experiment, for each target layer, parameters with smaller weight magnitudes are set to zero according to the preset pruning rate, and 10 epochs of fine-tuning are performed on the original dataset to recover the performance degradation caused by pruning.

Fig. 5 shows the distribution of normalized Hamming distances between the model and its pruned versions under different pruning rates. The experimental results indicate that with the HOS-based method, when the pruning rate is lower than 70$\%$, all normalized Hamming distances are significantly below the threshold, demonstrating that the method has good robustness against moderate-intensity pruning under the condition of non-Gaussian parameter distribution. In contrast, the normalized distances of the NTS-based method increase significantly when the pruning rate reaches 30$\%$, indicating that its discriminative ability is limited in this scenario.
\begin{figure}
\centerline{\includegraphics[width=21pc]{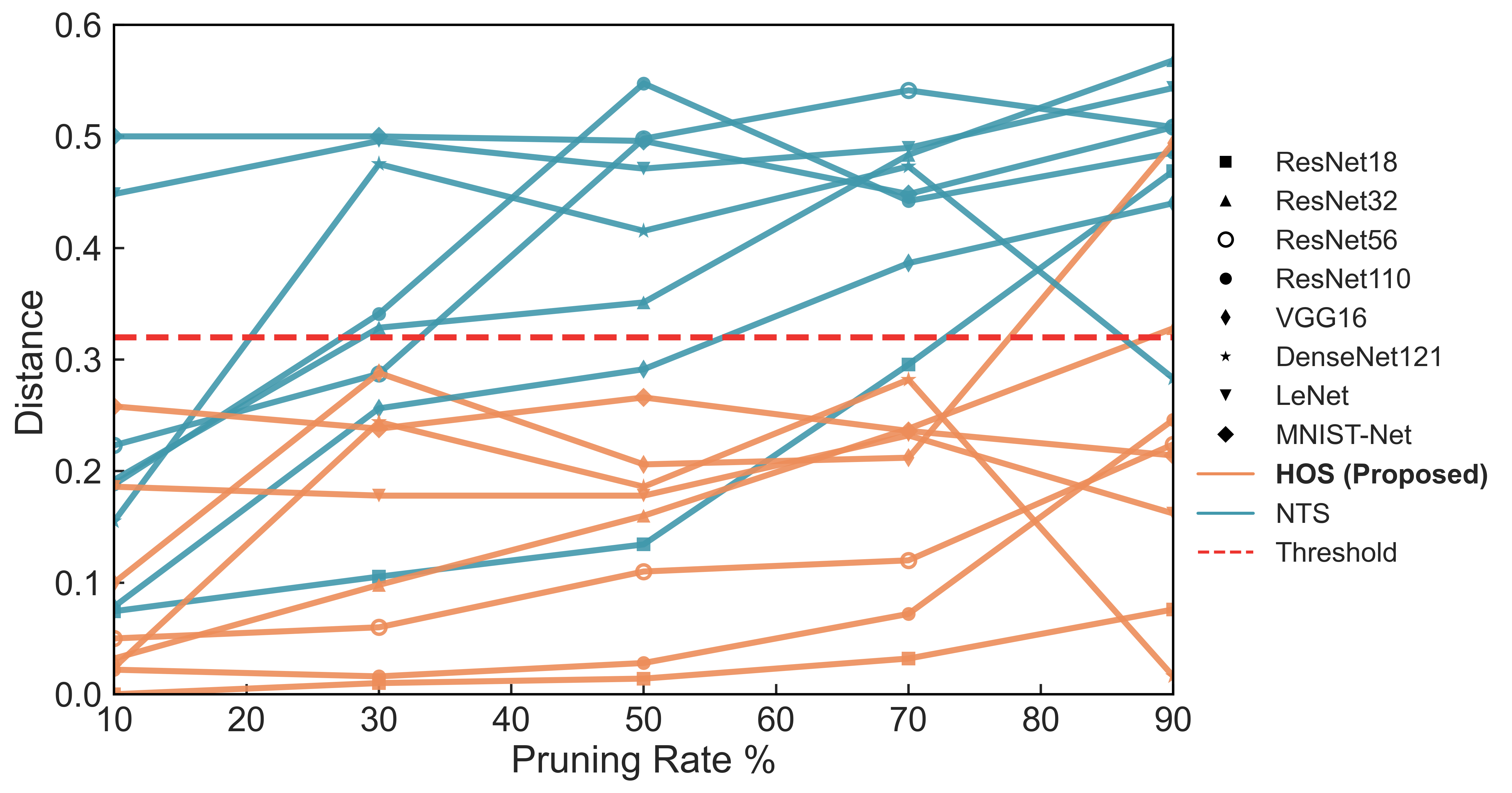}}
\caption{Normalized hamming distances of hash codes between the original models and the pruned versions using NTS and HOS methods.}
\end{figure}
\subsubsection{Knowledge Distillation}
It is assumed that adversaries employ knowledge distillation techniques to construct student models that approximate the functionality of teacher models, with the objective of evading hash-based detection. In this process, a student model sharing the same architecture as the teacher model is trained. The optimization is performed by simultaneously minimizing the discrepancy with the softened outputs of the teacher model, using a temperature parameter of 4.0, and incorporating the ground-truth labels with a mixing weight of 0.7.

Fig. 6 presents the normalized Hamming distances between the original model and its distilled versions under different training epochs. The results show that the distances obtained by the HOS-based method are always distinctly below the threshold, indicating that the hash representations maintain good stability even though distillation introduces large distribution differences. In contrast, the distances of the NTS-based method exceed the threshold in all epochs, which suggests that it is highly sensitive to the distillation process under the given settings and thus difficult to maintain reliable identification.

In summary, our HOS-based method not only maintains the discriminability for models with different structures and exhibits good robustness against fine-tuning, pruning, and distillation, but also avoids the reliance of the NTS-based method on the assumption of Gaussian distribution, thus being applicable to a wider range of scenarios.

\begin{figure}
\centerline{\includegraphics[width=21pc]{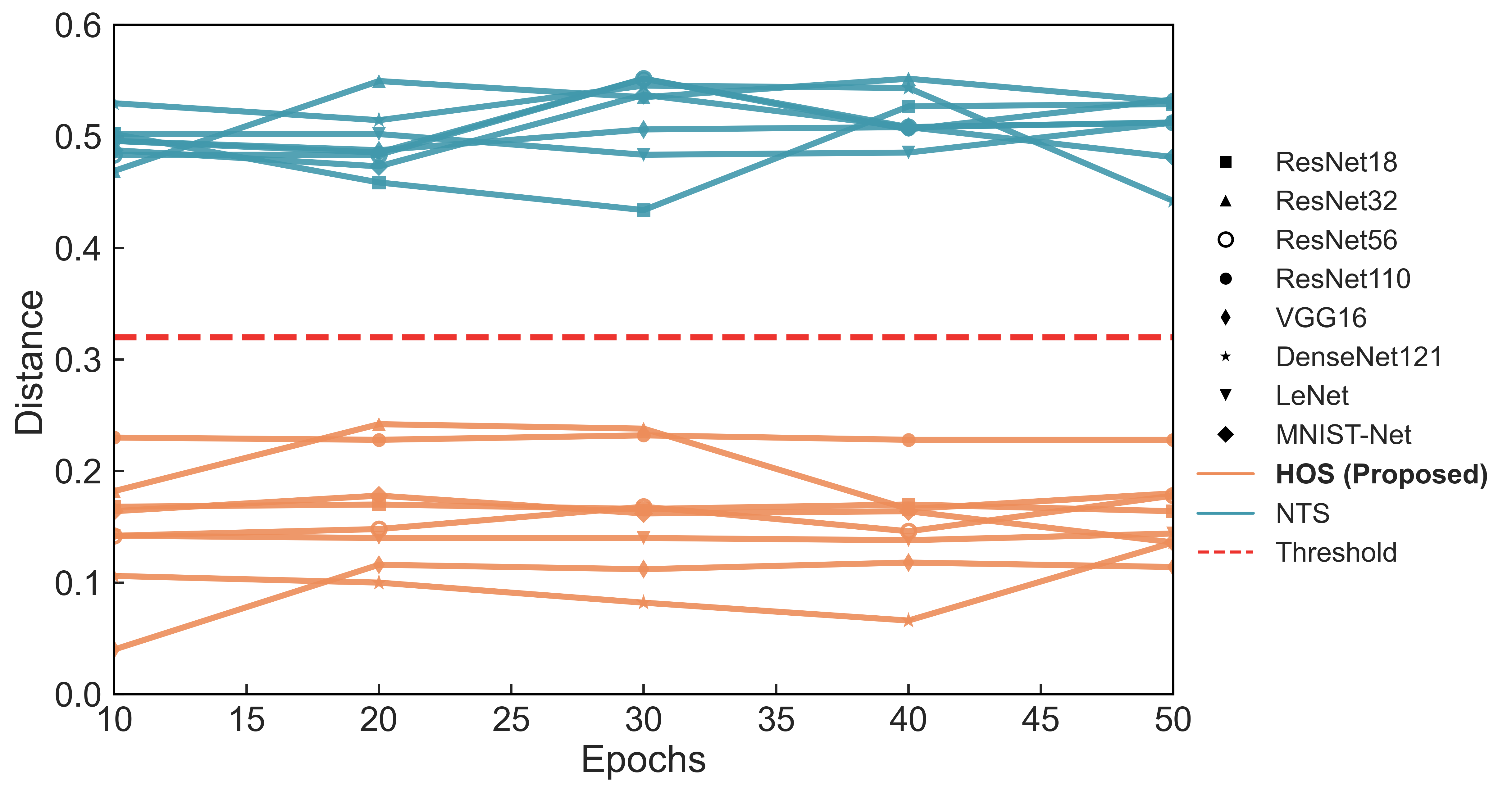}}
\caption{Normalized hamming distances of hash codes between the original models and the distilled versions using NTS and HOS methods.}
\end{figure}

\subsection{Tamper Detection}
To evaluate the necessity of integrating the chaotic mapping system into the proposed model tampering detection framework, we conducted comprehensive experiments on the eight representative models in Table I. Under predefined tampering rates, model parameters were randomly perturbed using Gaussian noise. The tampered models were subsequently examined by two detection schemes: (i) the proposed method incorporating the chaotic mapping system (denoted as Proposed), (ii) a baseline counterpart without the chaotic mapping system (denoted as Baseline). As described in Section III, parameter partitioning was first performed prior to detection. In particular, for the ResNet family, DenseNet121 and VGG16, parameters were divided into 450 blocks, whereas, due to their relatively smaller parameter scale, MNIST-Net and LeNet were partitioned into 100 blocks.

The detection results are shown in Tables \Rmnum{4}, \Rmnum{5}, and \Rmnum{6}. It is evident that, across all values of $\alpha$, the proposed chaotic method consistently achieved localization accuracy $R_t$ greater than 99\%. Such performance strongly indicates that this approach is capable of accurately localizing tampering under both weak and strong perturbations. By contrast, the baseline exhibited considerably inferior performance across all tampering rates. Notably, this deficiency became more pronounced when applied to complex architectures such as models in Tables \Rmnum{4} and \Rmnum{5}, as compared to simpler models such as MNIST-Net and LeNet in Table \Rmnum{6}. These findings highlight the indispensable role of the chaotic mapping system, and further substantiate its contribution to enhancing the robustness of the proposed tampering detection framework.

\begin{table}[H]
\centering
\caption{Statistics of tampering detection accuracy for ResNet models under malicious attacks. In this paper, $\alpha$ denotes the tampering ratio, $\sigma$=0.1 is the standard deviation of the Gaussian noise, and $R_t$ indicates the tamper recognition accuracy.}
\begin{tblr}{
  colspec = {c *{7}{c}}, 
  cell{1}{3} = {c=5}{},
  cell{2}{1} = {c=2}{},
  cell{3}{1} = {r=2}{},
  cell{5}{1} = {r=2}{},
  cell{7}{1} = {r=2}{},
  cell{9}{1} = {r=2}{},
  vline{3} = {3-10}{},
  hline{1-3,11} = {-}{},
}
   &        & Localization accuracy rate $R_t$ &     &     &     &     \\
$\alpha$ &  & 10\% & 20\% & 30\% & 40\% & 50\% \\
ResNet18 & Proposed & \textbf{100\%} & \textbf{100\%} & \textbf{100\%} & \textbf{100\%} & \textbf{100\%} \\
         & Baseline & 51.1\% & 48.9\% & 48.2\% & 47.8\% & 48.0\% \\
ResNet32 & Proposed & \textbf{100\%} & \textbf{100\%} & \textbf{100\%} & \textbf{100\%} & \textbf{100\%} \\
         & Baseline &  88.9\%   &  92.2\%   &  92.6\%   &  87.8\%   &  88.4\%   \\
ResNet56 & Proposed & \textbf{100\%} & \textbf{100\%} & \textbf{100\%} & \textbf{100\%} & \textbf{100\%} \\
         & Baseline &  80.0\%  &  84.4\%  &  85.2\%    & 86.1\%   &   86.2\%  \\
ResNet110& Proposed & \textbf{100\%} & \textbf{100\%} & \textbf{100\%} & \textbf{100\%} & \textbf{99.6\%} \\
         & Baseline &  77.8\%   &   78.9\%   &   77.0\%     & 75.0\%   &  76.0\%   
\end{tblr}
\end{table}
\begin{table}[H]
\centering
\caption{Statistics of tampering detection accuracy for DenseNet121 and VGG16 under malicious attacks.}
\begin{tblr}{
  colspec = {c *{7}{c}}, 
  cell{1}{3} = {c=5}{},
  cell{2}{1} = {c=2}{},
  cell{3}{1} = {r=2}{},
  cell{5}{1} = {r=2}{},
  vline{3} = {3-6}{},
  hline{1-3,7} = {-}{},
}
                          &         & Localization accuracy
  rate $R_t$ &        &        &        &        \\
$\alpha$ &         & 10\%                            & 20\%   & 30\%   & 40\%   & 50\%   \\
DenseNet121                  & Proposed & \textbf{100\%}                           & \textbf{100\%}  & \textbf{100\%}  & \textbf{100\%}  & \textbf{100\%}  \\
                          & Baseline & 53.3\%                          & 54.4\% & 49.6\% & 51.7\% & 53.3\% \\
VGG16                  & Proposed &      \textbf{100\%}                           &    \textbf{100\%}    &    \textbf{100\%}    &    \textbf{100\%}    &      \textbf{100\%}  \\
                          & Baseline &     33.3\%                            &  37.8\%      &    32.6\%    &   32.8\%     &   30.7\%     
\end{tblr}
\end{table}

\begin{table}
\centering
\caption{Statistics of tampering detection accuracy for MNIST-Net and LeNet under malicious attacks.}
\begin{tblr}{
  colspec = {c *{7}{c}}, 
  cell{1}{3} = {c=5}{},
  cell{2}{1} = {c=2}{},
  cell{3}{1} = {r=2}{},
  cell{5}{1} = {r=2}{},
  vline{3} = {3-6}{},
  hline{1-3,7} = {-}{},
}
                          &         & Localization accuracy
  rate $R_t$ &        &        &        &        \\
$\alpha$ &         & 10\%                            & 20\%   & 30\%   & 40\%   & 50\%   \\
MNIST-Net                  & Proposed & \textbf{100\%}                           & \textbf{100\%}  & \textbf{100\%}  & \textbf{100\%}  & \textbf{100\%}  \\
                          & Baseline & 80.0\%                          & 95.0\% & 86.7\% & 87.5\% & 82.0\% \\
LeNet                  & Proposed &     \textbf{100\%}                            &    \textbf{100\%}    &      \textbf{100\%}  &  \textbf{100\%}      & \textbf{100\%}       \\
                          & Baseline &       90.0\%                          &  90.0\%      &    86.7\%    &    80.0\%    &        86.0\%
\end{tblr}
\end{table}
\subsection{ Computational Cost}
The feasibility of the proposed scheme was validated by evaluating the computational overhead, i.e., the average execution time for hash generation. Specifically, the total time for extracting hashes was measured across 200 models, including 8 original models and 192 derivative models obtained through fine-tuning, pruning, and distillation. The average extraction time per hash was then recorded, and the qualitative comparison is given in Table \Rmnum{6}.

When implemented on a single AMD Ryzen 5 5600 6-Core CPU using the PyTorch framework, the proposed method achieved an average computational time of only 1.052 s. In particular, the weight selection stage and the feature generation stage required 1.007 s and 0.045 s on average, respectively. Compared with existing methods, our approach not only enables simultaneous model piracy identification and model tampering localization, but also avoids the reliance on DNNs that require substantial training time, thereby demonstrating its superiority.
\begin{table}[H]
\centering
\caption{Comparison of model hashing between the methods \cite{ref15}, \cite{ref16}, \cite{ref17} and the proposed method.}
\begin{tabular}{ccccc} 
\hline
                & Our Method & \cite{ref15} & \cite{ref16} & \cite{ref17}  \\ 
\hline
Lightweight Solution   & Yes          & Yes  & No  & No   \\
Model Piracy Identification      & Yes          & Yes  & Yes  & Yes   \\
Model Tampering Localization      & Yes          & No  & Yes  & Yes   \\
Applicability     & Yes          & No  & Yes  & Yes   \\
\hline
\end{tabular}
\end{table}
\section{Conclusion}
In this paper, we propose a lightweight CNN hashing method that integrates HOS features with the chaotic mapping mechanism. Without relying on additional neural network training or Gaussian distribution assumptions, this approach extracts skewness, kurtosis, and structural information to generate robust and discriminative hash codes. The chaotic mapping further amplifies subtle variations in parameters, enabling precise tamper localization.

Comprehensive experiments demonstrate that the proposed method exhibits excellent discriminative capability across different models and strong robustness against common modifications such as fine-tuning, pruning, and knowledge distillation. The tampering detection mechanism can accurately identify and localize modified parameter blocks. Compared to existing methods, our technique offers greater adaptability and lower computational cost, making it highly suitable for practical model copyright protection and integrity verification scenarios.

\begin{IEEEbiography}[{\includegraphics[width=1in,height=1.25in,clip,keepaspectratio]{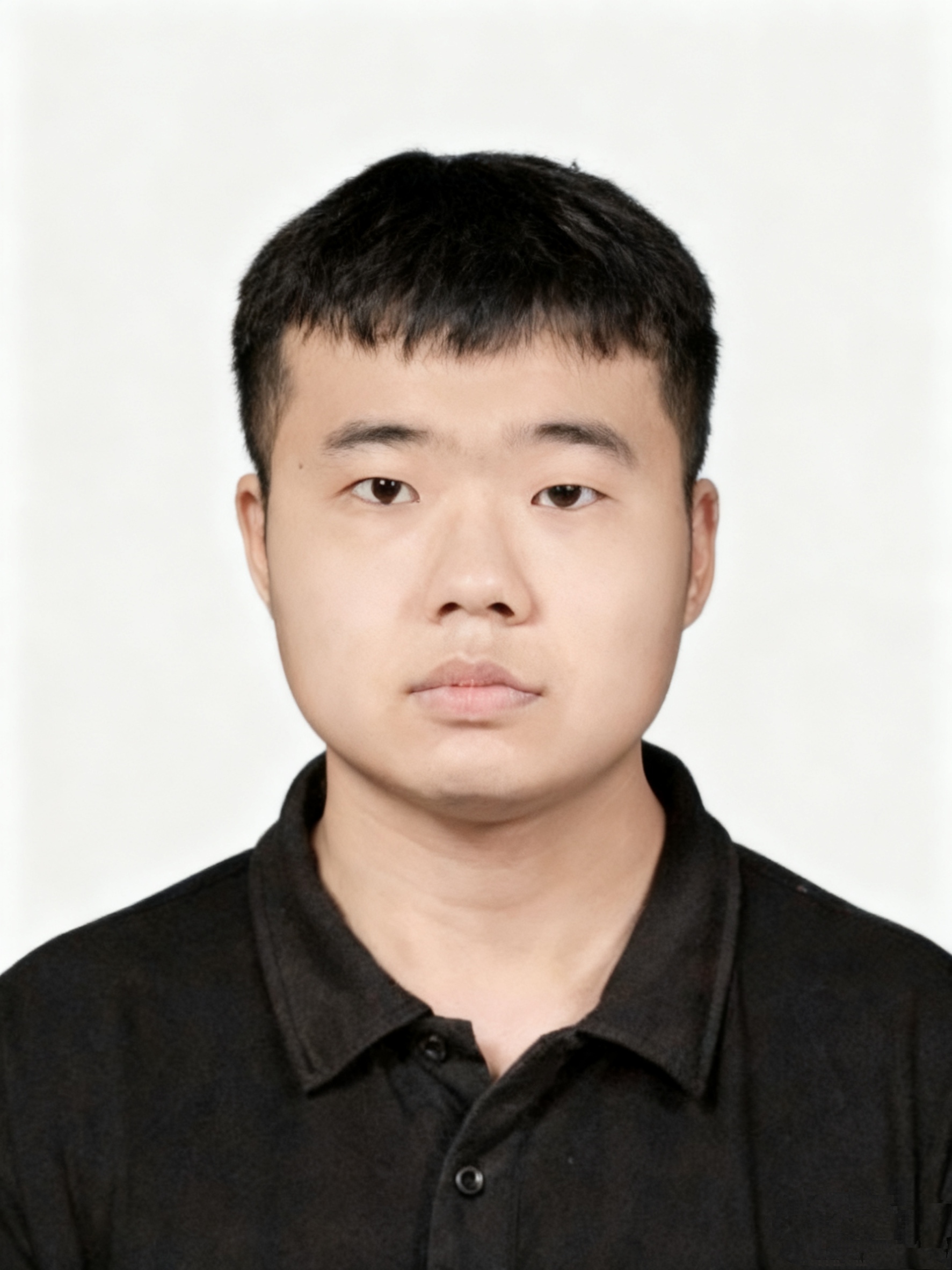}}]{Kunming Yang}
received the B.S. degree from the School of Electronic and Information Engineering, Southwest University, Chongqing, China, in 2024. He is currently pursuing the M.S. degree with the School of Electronic and Information Engineering, Southwest University, Chongqing. His current research interests include and neural network model protection memristor-based neural networks. 
\end{IEEEbiography}

\begin{IEEEbiography}[{\includegraphics[width=1in,height=1.25in,clip,keepaspectratio]{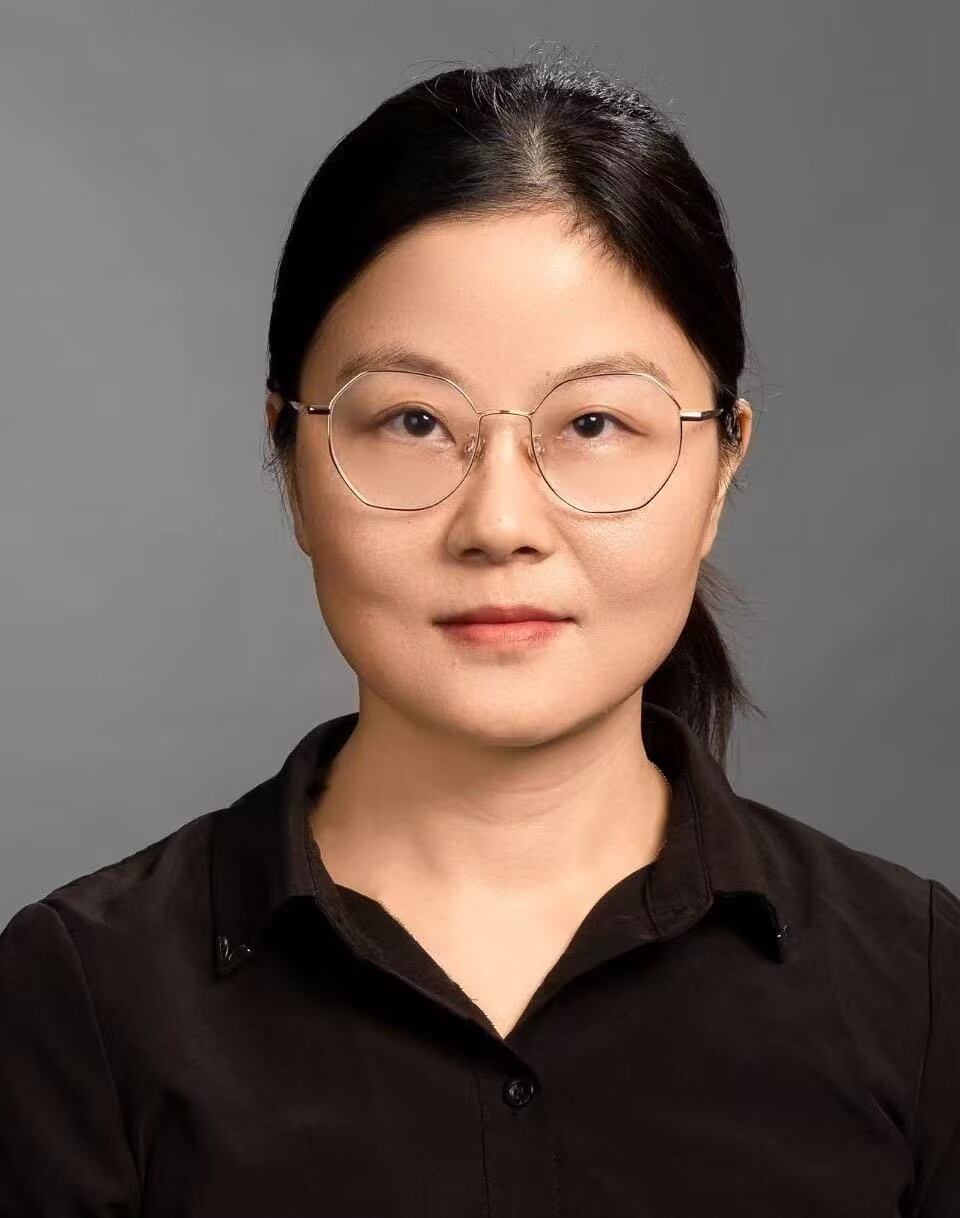}}]{Ling Chen}
received the Ph.D. degree in computer science and technology from Chongqing University, Chongqing, China, in 2014. From 2013 to 2014, she was a joint Ph.D. Student of electrical engineering at the University of Pittsburgh. She has been an Associate Professor at the School of Electronic and Information Engineering, Southwest University, since July 2019. She has published over 30 papers in international journals and conferences and has been granted several national invention patents. Her research interests include brain-like neural networks, memristors, image processing and pattern recognition, and nonlinear systems.
\end{IEEEbiography}

\section{References}

\begin{thebibliography}{99}
\setcounter{enumiv}{0}

\bibitem{ref1} S. Mehta, V. Kukreja and R. Gupta, ``Empowering Precision Agriculture: Detecting Apple Leaf Diseases and Severity Levels with Federated Learning CNN,'' in {\em International Conference on Intelligent Technologies (CONIT)}, 2023, pp. 1–6. 

\bibitem{ref2} M. Peng, Y. Liu, I. A. Qadri, U. A. Bhatti, B. Ahmed, N. M. Sarhan, and E. M. Awwad, ``Advanced image segmentation for precision agriculture using CNN-GAT fusion and fuzzy C-means clustering,'' {\em Comput. Electron. Agricult.}, vol. 226, Nov. 2024, Art. no. 109431.

\bibitem{ref3} Asadi, R., Queguineur, A., Wiikinkoski, O., Mokhtarian, H., Aihkisalo, 
T., Revuelta, A. and Ituarte, I.F., ``Process monitoring by deep neural networks in directed energy deposition: Cnn-based detection, segmentation, and statistical analysis of melt pools,'' {\em Rob. Comput. Integr. Manuf.}, to be published. vol. 87, 2024, 102710.

\bibitem{ref4} J. Aldrini, I. Chihi, and L. Sidhom, ``Fault diagnosis and self-healing for smart manufacturing: a review,'' {\em J. Intell. Manuf.}, vol. 35, no. 6, pp. 2441–2473, Aug. 2024.

\bibitem{ref5} C. Liu, H. Zhu, D. Tang, Q. Nie, S. Li, Y. Zhang, and X. Liu, ``A transfer learning CNN-LSTM network-based production progress prediction approach in IIoT-enabled manufacturing,'' {\em Int. J. Prod. Res.}, vol. 61, no. 12, pp. 4045–4068, Jun. 2023.

\bibitem{ref6} D. Hupkes, M. Giulianelli, V. Dankers, M. Artetxe, Y. Elazar, T. Pimentel, C. Christodoulopoulos, K. Lasri, N. Saphra, A. Sinclair, D. Ulmer, F. Schottmann, K. Batsuren, K. Sun, K. Sinha, L. Khalatbari, M. Ryskina, R. Frieske, R. Cotterell, and Z. Jin, ``A taxonomy and review
 of generalization research in NLP,'' {\em Nature Mach. Intell.}, vol. 5, no. 10, pp. 1161–1174, Oct. 2023. 

\bibitem{ref7} J. R. Jim, M. A. R. Talukder, P. Malakar, M. M. Kabir, K. Nur, and M. F. Mridha, ``Recent advancements and challenges of NLP-based sentiment analysis: A state-of-the-art review,'' {\em Natural Lang. Process. J.}, vol. 6, Mar. 2024, Art. no. 100059.

\bibitem{ref8} Y. Uchida, Y. Nagai, S. Sakazawa, and S. Satoh, ``Embedding watermarks into deep neural networks,'' in {\em Proc. ACM Int. Conf. Multimedia Retrieval}, Bucharest, Romania, 2017, pp. 269–277.

\bibitem{ref9} X. Zhao, Y. Yao, H. Wu, and X. Zhang, ``Structural watermark
ing to deep neural networks via network channel pruning,'' in {\em Proc. IEEE Int. Workshop Inf. Forensics Secur. (WIFS)}, Dec. 2021, pp. 1–6.

\bibitem{ref10} J. Zhao, Q. Hu, G. Liu, X. Ma, F. Chen, and M. M. Hassan, ``AFA: Adversarial fingerprinting authentication for deep neural networks,'' {\em Comput. Commun.}, vol. 150, pp. 488–497, Jan. 2020.

\bibitem{ref11} Y. Li, Z. Zhang, B. Liu, Z. Yang, and Y. Liu, ``ModelDiff: Testing-based
 DNNsimilaritycomparisonformodelreusedetection,'' in {\em Proc.30th ACM SIGSOFT Int. Symp. Softw. Testing Anal.}, 2021, pp. 139–151.

\bibitem{ref12} S. M. Abdullahi, H. Wang, and T. Li, ``‘Fractal coding-based robust and alignment-free fingerprint image hashing,'' {\em IEEE Trans. Inf. Forensics Security}, vol. 15, pp. 2587–2601, 2020.

\bibitem{ref13} X. Liang, Z. Tang, Z. Huang, X. Zhang, and S. Zhang, ``Efficient hashing method using 2D-2D PCA for image copy detection,'' {\em IEEE Trans. Knowl. Data Eng.}, vol. 35, no. 4, pp. 3765–3778, Apr. 2023.

\bibitem{ref14} Y. Shen, L. Liu, F. Shen, and L. Shao, ``Zero-shot sketch-image hashing,'' in {\em Proc. IEEE Conf. Comput. Vis. Pattern Recognit.}, Mar. 2018, pp. 3598–3607.

\bibitem{ref15} H. Chen, H. Zhou, J. Zhang, D. Chen, W. Zhang, K. Chen, G. Hua, and N. Yu, ``Perceptual hashing of deep convolutional neural networks for model copy detection,'' {\em ACM Trans. Multimedia Comput., Commun., Appl.}, vol. 19, no. 3, pp. 1–20, 2023.

\bibitem{ref16} C. Xiong, G. Feng, X. Li, X. Zhang, and C. Qin, ``Neural network model protection with piracy identification and tampering localization capability,'' in {\em Proc. 30th ACM Int. Conf. Multimedia}, Oct. 2022, pp. 2881–2889.

\bibitem{ref17} R. Liu, H. Chen, B. Zhao, K. Chen, and W. Zhang, ``Graph-Embedded Structure-Aware Perceptual Hashing for Neural Network Protection and Piracy Detection,'' in {\em Proc. IEEE Conf. Comput. Vis. Pattern Recognit. (CVPR)}, Jun. 2025, pp. 20169-20178.

\bibitem{ref18} C. Blundell, J. Cornebise, K. Kavukcuoglu, and D. Wierstra, ``Weight uncertainty in neural network,'' in {\em Proc. Int. Conf. Mach. Learn.}, 2015, pp. 1613–1622.

\bibitem{ref19} C. Chatzikonstantinou, G. T. Papadopoulos, K. Dimitropoulos, and P. Daras, ``Neural network compression using higher-order statistics and auxiliary reconstruction losses,'' in {\em  2020 IEEE/CVF Conference on Computer Vision and Pattern Recognition Workshops (CVPRW)}, pp. 3077--3086, 2020.

\bibitem{ref20} A. Barreira, D. Nelson, A. Pillepich, V. Springel, F. Schmidt, R. Pakmor, L. Hernquist, and M. Vogelsberger, ``Separate Universe simulations with IllustrisTNG: baryonic effects on power spectrum responses and higher-order statistics,'' {\em Mon. Not. R. Astron. Soc.}, vol. 488, no. 2, pp. 2079--2092, 2019.

\bibitem{ref21} S. A. Khoshnevis and R. Sankar, ``Diagnosis of Parkinson’s disease using
 higher order statistical analysis of alpha and beta rhythms,'' {\em Biomed. Signal Process. Control}, vol. 77, Aug. 2022, Art. no. 103743.

\bibitem{ref22} H. Li, Z. Hua, H. Bao, L. Zhu, M. Chen, and B. Bao, ``Two-dimensional memristive hyperchaotic maps and application in secure communication,'' {\em IEEE Trans. Ind. Electron.}, vol. 68, no. 10, pp. 9931–9940, Oct. 2021.

\bibitem{cifar} A. Krizhevsky, and G. Hinton, ``Learning multiple layers of features from tiny images,'' 2009.

\bibitem{MNIST} D. Ciresan, U. Meier, and J. Schmidhuber, ``Multi-column deep neural networks for image classification,'' in {\em Proc. IEEE Conf. Comput. Vis. Pattern Recognit.}, Jun. 2012, pp. 3642-3649.

\bibitem{densenet} G. Huang, Z. Liu, L. Van Der Maaten, and K. Q. Weinberger, ``Densely connected convolutional networks,'' in {\em Proc. IEEE Conf. Comput. Vis. Pattern Recognit. (CVPR)}, Jul. 2017, pp. 2261–2269.

\bibitem{res18-32-56-110} K. He, X. Zhang, S. Ren, and J. Sun, ``Deep residual learning for image recognition,'' in {\em Proc. IEEE Conf. Comput. Vis. Pattern Recognit. (CVPR)}, Jun. 2016, pp. 770–778.

\bibitem{vgg16} K. Simonyan, and A. Zisserman, ``Very deep convolutional networks for large-scale image recognition,'' in {\em Proc. Int. Conf. Learn. Representations}, 2015, pp. 1–14.

\bibitem{mnist-net} N. Carlini, and D. Wagner, ``Towards evaluating the robustness of neural networks,'' in {\em Proc. IEEE Symp. Secur. Privacy}, 2017, pp. 39–57.

\bibitem{lenet} Y. Lecun, L. Bottou, Y. Bengio, and P. Haffnerr, ``Gradient-based learning applied to document recognition,'' {\em Proc. IEEE}, vol. 86, no. 11, pp. 2278–2324, Nov. 1998.
\end{thebibliography}
\end{document}